\documentclass[12pt]{article}

\usepackage{amsmath,amssymb}
\usepackage{cite}
\usepackage{textcmds}
\usepackage[unicode,linktocpage=true,plainpages=false,pdfpagelabels=false]{hyperref}

\newcommand{\dd}{\partial}
\newcommand{\de}{\delta}
\newcommand{\m}{\mu}
\newcommand{\n}{\nu}
\newcommand{\ls}{\left(}
\newcommand{\rs}{\right)}
\newcommand{\na}{\nabla\!}
\newcommand{\al}{\alpha}
\newcommand{\be}{\beta}
\newcommand{\ka}{\varkappa}
\newcommand{\te}{\theta}
\newcommand{\ta}{\tau}
\newcommand{\la}{\lambda}
\newcommand{\La}{\Lambda}
\newcommand{\ga}{\gamma}
\newcommand{\str}[1]{\mathrel{\mathop{\longrightarrow}\limits_{#1}}}
\newcommand{\om}{\omega}

\newcommand{\disn}[2]{$$\displaylines{\refstepcounter{equation}%
            \label{#1}\hskip 1em minus 1em #2\hfilneg}$$}
\newcommand{\nom}{\hfil\hskip 1em minus 1em (\theequation)}
\newcommand{\ns}{\hfill\cr\hfill}

\begin{document}

\title{Classical electromagnetic potential as a part\\
of gravitational connection: ideas and history}

\author{
N.~V.~Kharuk\thanks{E-mail: natakharuk@mail.ru},
S.~A.~Paston\thanks{E-mail: s.paston@spbu.ru},
A.~A.~Sheykin\thanks{E-mail: a.sheykin@spbu.ru}\\
{\it Saint Petersburg State University, Saint Petersburg, Russia}
}

\date{\vskip 15mm}

\maketitle

\begin{abstract}
We consider a natural form of unified theory of gravity and electromagnetism which was somehow missed
at the time of intense search for such an unification, and was noticed only in 1978 but remained quite unknown.
The basic idea of this unification is to use the metric and non-symmetric connection as independent variables,
which generalizes the so-called Palatini formalism. The certain components of connection in the appearing theory
can be naturally identified with electromagnetic potential, and  with the proper choice of action
the Einstein-Maxwell equations are reproducing. In this paper we compare such an approach with the known
ideas of unification.
Also we propose the more consistent way of including matter
(in the form of classical particles) in the theory
and briefly discuss the perspectives of further development of this approach.
\end{abstract}

\newpage

\section{Introduction}\label{vved}
A distinctive feature of the gravity description in the framework of general relativity is a variety of its formulations, mostly geometric ones. Among the geometric formulations there are not only purely metric ones (as in the original General Relativity (GR)), but also others, which possess the existence of additional geometric structures that are not reducible to the metric. In many cases for the introduction of such structures one need to extend the formalism or even the physical essence of GR, e.~g., for the introduction of an orthogonal frame \cite{cartan2001} or an isometric embedding \cite{regge,statja18}.

However, there is a geometrical object which is \textit{a priori} not reducible to metric: a connection on the considered manifold.
In the original Einsteinian formulation the connection is present, but it is postulatively expressed through metric. One can thus raise a natural question: if one rejects the initial consistency of connection with metric, what the resulting theory would be like?

It is now well known that if the metric and connection are assumed to be independent variables, then in the assumption of the zero torsion the variation of the Einstein-Hilbert action with respect to metric leads to usual Einstein equations, and with respect to connection --- the expression for connection in the form of Christoffel symbols which are symmetric. The resulting theory is equivalent to GR. In the literature it is usually called the Palatini formalism (or Hilbert-Palatini formalism), although such a naming is not quite correct. In the paper by Palatini \cite{palatini} the variational principle for Einstein-Hilbert (EH) action was considered; but connection was not treated as an independent variable (a detailed historical survey of this question can be found in \cite{ferraris_pal}). For the first time  it seems to have been done by Einstein \cite{ein_conn} in the attempt of unified field theory (UFT) construction. Note that his approach was more general than Palatini one: he assumed the \textit{ metric} is non-symmetric as well as the connection; and identified the antisymmetric part of the metric with the electromagnetic (EM) field tensor.

In the present paper we want to discuss a variant of gravity and electromagnetism unification which was somehow missed by Einstein and which seems far more natural. It appears when one assumes the independency of metric and connection, considering metric as symmetric and connection as non-symmetric, i.~e. all 64 components of connection is considered to be independent. Then it turns out to be possible to identify some components of the connection with the EM potential. During the XX century researchers closely approached this idea more than once (we give a short historical review of the corresponding publications in the section \ref{istor}), but for the first time it was fully embodied only in 1978 in the paper \cite{Krechet}. Moreover, the paper \cite{Krechet} remains practically unknown to the scientific community because of poor availability of the journal in which it was published. The main addition of our work to the previously known results is the natural method of coupling the connection with matter (in the form of a set of relativistic point particles), after which the theory takes the form of usual Einstein-Maxwell theory with matter.

In the section \ref{urav} we derive the field equations for the metric and connection using the variational principle for the standard EH action. This derivation is given here for ease of reading: although this interesting result was repeatedly obtained earlier by many authors, it is still not well known.
In the section \ref{mater} we consider a problem of including the matter (in the form of relativistic particles) in the variational principle. As we will show below, it is possible to organize non-minimal (i.~e. not reducible to the usual covariantization of derivatives) coupling of matter with connection that does not broke the general covariance. The field equations of the resulting theory are examined in the section~\ref{urdv}. The analysis shows that they reproduce Einstein-Maxwell equations.

The simplicity and naturalness of the constructed theory raise a question: why it had not been discovered by any of researchers who work in the field of the UFT: Einstein, Weyl, Eddindton etc.? The section~\ref{istor} is devoted to the review of early theories of geometrized electrodynamics and some recent works which are related
to the theory discussed here. In the last section~\ref{zakl} we briefly discuss the problems and perspectives of generalization of this theory.

\section{The affine-metric formulation of pure gravity}\label{urav}
Let us consider the affine-metric theory of gravity, assuming that the metric and connection are independent. The metric, as usual, is supposed to be symmetric, whereas connection is not (i.~e. the torsion is nonzero). In the present paper we choose the $(+,-,-,-)$ signature of spacetime. As an action of pure gravity (without matter) we choose the usual EH one
\begin{align}\label{1}
	S_1=-\frac{1}{2\ka}\int\! d^4 x\, \sqrt{-g}\,g^{\n\be}\, R^{\m}\,\!_{\n\m\be},
\end{align}
where the Riemann-Christoffel curvature tensor is expressed only through independent connection $\Gamma^\alpha_{\mu\nu}$:
\begin{align}\label{s1}
	R^{\mu}\,\!_{\nu\alpha\beta} = \partial_\alpha \Gamma^{\mu}_{\beta\nu}-\partial_\be \Gamma^{\mu}_{\al\nu}+
 \Gamma^{\mu}_{\alpha\gamma} \Gamma^{\gamma}_{\beta\nu}-\Gamma^{\mu}_{\be\gamma} \Gamma^{\gamma}_{\al\nu}.
\end{align}
In \eqref{1} $g^{\n\be}$ is a metric, $g$ is its determinant and $\ka$ is an Einstein gravitational constant.
To fix the order of the indices of connection (which is not symmetric), we define an action of covariant derivative on an arbitrary vector as follows:
\disn{s2.0}{
D_\m u^\al=\dd_\m u^\al+\Gamma^\alpha_{\mu\nu}u^\n,
\nom}
i.~e. the first index of connection is the one that is related to the derivative. Note that while the connection remains independent, the Ricci tensor
\disn{s2.1}{
R_{\n\be}=R^{\m}\,\!_{\n\m\be}
\nom}
is not symmetric.

One can notice that the action \eqref{1} possesses not only diffeomorphic invariance, but also an additional invariance with respect to the so-called \qq{projective transformations} \cite{sandberg1975} (it is worth noting that such transformations were considered earlier by Einstein in \cite{ein1955-1})
\begin{align}\label{gauge}
	{\Gamma'}^\alpha_{\mu\nu} = {\Gamma}^\alpha_{\mu\nu} + f_\mu \delta^\alpha_\nu
\end{align}
with the arbitrary function $f_\m$, i.~e. some gauge transformations.
Indeed, as one can see from \eqref{s1}, such a transformation adds an explicitly antisymmetric expression to the Ricci tensor
\begin{align}\label{ss1}
	{R'}_{\n\be} = {R}_{\n\be} + \partial_\n f_\be - \partial_\be f_\n
\end{align}
(this result was obtained in \cite{ein1955-1}) which in \eqref{1} is contracted with symmetric metric and thus does not give a contribution. The action is a quadratic function of connection, therefore its symmetry with respect to transformations \eqref{gauge} leads to the fact that quadratic form which is contained in it is degenerate and has rank 60 while its dimension is 64 (64 is the number of independent components of connection  $\Gamma^\alpha_{\mu\nu}$). Such a degeneracy of the action means that one can not find all the components of connection from the equation which is obtained from the action by varying it with respect to connection (note that in Palatini formalism with symmetric connection it is possible). A part of connection components must remain arbitrary. Let us prove it by a direct calculation.

Let us find the variation of the action \eqref{1} with respect to the connection. Using \eqref{s1} one can easily find that
\disn{s2}{
\de S_1=-\frac{1}{2\ka}\int\! d^4 x\, \sqrt{-g}\,g^{\n\be}\,
\ls D_\m \de\Gamma^{\mu}_{\beta\nu}-D_\be \de\Gamma^{\mu}_{\m\nu}+S^\ga{}_{\m\be}\de\Gamma^{\mu}_{\ga\nu}\rs,
\nom}
where a notation for the torsion was used:
\disn{s3}{
S^\al{}_{\m\n}=\Gamma^\al_{\m\n}-\Gamma^\al_{\n\m}.
\nom}
Let us introduce the quantity which determines the difference between the connection and its Riemannian part:
\disn{s4}{
C_\m{}^\al{}_\n=\Gamma^\al_{\m\n}-\bar\Gamma^\al_{\m\n},
\nom}
where $\bar\Gamma^\al_{\m\n}$ is a symmetric Riemannian connection expressed through metric (Christoffel symbols):
\begin{align}\label{s5}
\bar\Gamma^\alpha_{\mu\nu} = \frac{1}{2} g^{\alpha\beta}(\partial_\mu g_{\nu\beta}+\partial_\nu g_{\mu\beta}-\partial_\beta g_{\mu\nu}).
\end{align}
A so-called non-metricity can be easily expressed through $C_\m{}^\al{}_\n$
\disn{s6}{
D_\m g^{\n\be}=C_\m{}^{\n\be}+C_\m{}^{\be\n},
\nom}
as well as the torsion
\disn{s9.1}{
S^\al{}_{\m\n}=C_\m{}^\al{}_\n-C_\n{}^\al{}_\m.
\nom}

Note that using \eqref{s4} one can (up to a boundary term) write a relation

\disn{s7}{
\int\! d^4 x\, \sqrt{-g}\, D_\m u^\m=
\int\! d^4 x\, \sqrt{-g}\, \ls \bar D_\m u^\m+C_\m{}^\m{}_\n  u^\n\rs=
\int\! d^4 x\, \sqrt{-g}\, C_\m{}^\m{}_\n  u^\n,
\nom}
for the arbitrary vector $u^\m$, where $\bar D_\m$ is a Riemannian covariant derivative which contains only the Riemannian connection\eqref{s5}. Using this relation in the integration by parts as well as the formulas \eqref{s6} and \eqref{s9.1} one can perform a simple calculation and, dropping the boundary term, rewrite the variation \eqref{s2} in the following form:
\disn{s8}{
\de S_1=\frac{1}{2\ka}\int\! d^4 x\, \sqrt{-g}\ls
\ls C_\m{}^{\n\be}+C^{\n\be}{}_\m-C_\ga{}^\ga{}_\m g^{\n\be}\rs\de\Gamma^\m_{\be\n}-
C_\m{}^{\n\m}\de\Gamma^\be_{\be\n}\rs.
\nom}
The corresponding field equation is easy to obtain from it:
\disn{s9}{
C^{\m\n\be}+C^{\n\be\m}-C_\ga{}^{\ga\m} g^{\n\be}-C_\al{}^{\n\al}g^{\be\m}=0.
\nom}
Contracting the indices in multiple ways, one can easily notice that one way leads to the identity, whereas two others --- to the relations between different contractions of $C^{\m\n\be}$:
\disn{s10}{
C_{\ga\m}{}^{\ga}=C_\ga{}^\ga{}_\m=\frac{1}{4}C_{\m\ga}{}^\ga.
\nom}
However, the value of this contractions is not determined by \eqref{s9}, and remains arbitrary, as we mentioned above. Let us introduce the notation for it: $C_{\ga\m}{}^{\ga}\equiv \om A_\m$,
where $A_\m$ is an arbitrary vector field and $\om$ is a constant which will be fixed later.
Other components of $C^{\m\n\be}$ can be uniquely expressed through this quantity. To prove it, one need to consider a linear combination of three copies of  \eqref{s9}, which differ by cyclic permutation of indices. As a result, one obtains
\disn{s11}{
C_\m{}^\n{}_\be=\om A_\m\de^\n_\be,
\nom}
and the final expression for the connection can be found using \eqref{s4}:
\disn{s12}{
\Gamma^\al_{\m\n}=\bar\Gamma^\al_{\m\n}+\om A_\m\de^\al_\n.
\nom}

As can be seen, field equation corresponding to variation of the action \eqref{1} with respect to connection leads to the fact that the connection coincides with the Christoffel symbols \eqref{s5} up to the gauge transformation \eqref{gauge}. Therefore some of components of connection $\Gamma^\al_{\m\n}$ corresponding to vector $A_\m$ can not been expressed through metric, remaining arbitrary. For the connection \eqref{s12} both the torsion and non-metricity are expressed through this vector:
\disn{s13}{
S^\al{}_{\m\n}=\om\ls A_\m\de^\al_\n-A_\n\de^\al_\m\rs,\qquad
D_\m g^{\n\be}=2\om A_\m g^{\n\be}.
\nom}
Spacetime with such non-metricity is called the Weyl-Cartan spacetime, and vector $-2\om A_\m$ is called the Weyl vector \cite{hehl}.
Note that for the connection of the form \eqref{s12}, i.~e. on-shell, $A_\m$ is simply connected with the traces of torsion an non-metricity tensors:
\disn{s14}{
A^\m=\frac{1}{3\om}S^{\ga\m}{}_\ga=\frac{1}{2\om}D_\ga g^{\ga\m},
\nom}
whereas in the curvature tensor \eqref{s1} is is contained as follows:
\disn{s14.1}{
R^\m{}_{\n\al\be}=\bar R^\m{}_{\n\al\be}+\om\de^\m_\n\ls\partial_\al A_\be - \partial_\be A_\al\rs,
\nom}
where $\bar R^\m{}_{\n\al\be}$ is a Riemannian expression for the curvature tensor which is constructed from \eqref{s5}.

Now let us find the variation of \eqref{1} with respect to metric. This task turns out to be quite simple since the action depends on the metric itself but not on its derivatives. After the variation we obtain
\begin{align}\label{ein}
R_{\mu\nu}+R_{\nu\mu}-R\, g_{\mu\nu}  = 0,
\end{align}
where $R=g^{\al\be}R_{\al\be}$ is a scalar curvature.
The full system of field equations for the action \eqref{1} is a set of equations \eqref{ein} and  \eqref{s12}.
If \eqref{s12} is satisfied, then it is easy to obtain from \eqref{s14.1} that
\disn{s15}{
R_{\mu\nu} = \bar R_{\mu\nu} + \om\ls\partial_\m A_\n - \partial_\n A_\m\rs,
\nom}
where $\bar R_{\mu\nu}$ is a  Riemannian expression for the curvature tensor which is constructed from the Riemannian connection $\bar\Gamma^\al_{\m\n}$; therefore $\bar R_{\mu\nu}$ is symmetric and contains only metric. Making use of this fact, one can rewrite the equation \eqref{ein} in the usual form of the vacuum Einstein equations:
\disn{s16}{
\bar G_{\mu\nu}=0,
\nom}
where the Riemannian Einstein tensor $\bar G_{\mu\nu}$ is expressed through metric in a usual manner.

Thus for affine-metric formulation of pure gravity we see the satisfaction of vacuum Einstein equations for the metric, whereas the connection turns out to have its Riemannian value up to the gauge transformation \eqref{gauge} which leaves some degrees of freedom (DoF) of connection to be arbitrary. Such a theory can be treated \cite{sandberg1975} as equivalent to pure Einstein gravity if the vector field $A_\m$ is considered as a pure gauge DoF.

\section{ Interaction with relativistic particles}\label{mater}
Let us add the matter to the theory in the form of a set of relativistic point particles. First of all, let us add to the full action a standard term for relativistic particles with worldlines $x^\m_j(\ta)$ (index $j$ counts particles, $\ta$ parametrizes points of its worldlines) in a gravitational field defined by metric $g_{\m\n}$:
\begin{align}\label{s17}
S_2 = -\sum_j m_j \int\! d\ta\, \sqrt{\dot{x}_j^{\mu}(\ta)\dot{x}_j^\nu(\ta)g_{\mu\nu}(x_j(\ta))}.
\end{align}
Here $m_j$ is a mass of corresponding particle and $\dot{a}\equiv da/d\tau$.

If we restrict ourselves to a sum of \eqref{1} and \eqref{s17} in the action, then the forced acting on particles are equivalent to those in GR, and degrees of freedom corresponding to $A_\m$ (see \eqref{s12}) remain purely gauge. However, one can raise a question: what else local contributions containing gravitational and matter degrees of freedom one can add to the action assuming their invariance with respect to diffeomorphisms $x^\m\to x'^\m(x)$ as well as the reparametrizations of worldlines $\ta\to \ta'(\ta)$? It turns out that there is several possibilities to construct such contributions; one of them is the following:
\begin{align}\label{s18}
S_3 = -\sum_j \frac{q_j}{4\om} \int\! d\ta\, \dot{x}_j^{\be}(\ta) \Gamma^\m_{\be\m}(x_j(\ta)),
\end{align}
where $q_j$ are some arbitrary constants and the multiplier $1/(4\om)$ is written for the convenience of subsequent discussion. It is worth noting that in the case of Riemannian connection such a contribution is reduced to the addition of a full derivative with respect to $\tau$ to the Lagrangian due to the known feature of Riemannian connection:
\disn{s31}{
\bar\Gamma^\n_{\m\n}=\dd_\m\ln\sqrt{-g}.
\nom}
The choice of matter coupling with connection in the form \eqref{s18} leads to the most interesting results. The invariance of \eqref{s18} with respect to reparametrizations $\ta\to \ta'(\ta)$ is obvious, whereas its diffeomorphic invariance is present only up to a certain boundary terms and requires additional discussion.

Note that under the diffeomorphisms $x^\m\to x'^\m(x)$ the contracted connection transforms as
\begin{align}\label{part_conn}
{\Gamma}'^\m_{\be\m} = \frac{\partial x^\nu}{\partial x'^\be} \left( \Gamma^\m_{\nu\m} - \partial_\nu\ln\det\left| \frac{\partial x'}{\partial x} \right| \right),
\end{align}
so the quantity \eqref{s18} takes an increment
\begin{align}\label{diff_surf}
\Delta S_3 =
\sum_j \frac{q_j}{4\om} \int\! d\ta\, \dot{x}_j^{\be}(\ta) \partial_\be\ln\det\left| \frac{\partial x'}{\partial x} \right|=
\sum_j \frac{q_j}{4\om} \int\! d\ta\, \frac{d}{d\ta} \ln\det\left| \frac{\partial x'}{\partial x} \right|.
\end{align}
Such an increment in the form of the integral over full derivative does not affect the field equations of the theory. Its appearing after diffeomorphic transformations thus does not lead to breaking of general covariance of the field equations, as we explicitly show below.
Strictly speaking, the contribution \eqref{s18} to the action is not fully diffeomorphic invariant. It is invariant only with respect to a narrower group of coordinate transformations which are restricted by a condition
\disn{s19}{
\det\left| \frac{\partial x'}{\partial x} \right| \str{x^0\to\pm\infty} 1.
\nom}

It is worth noting that if the term \eqref{s18} is added to the full action, it is no longer invariant with respect to \eqref{gauge}. The invariance with respect to narrower group of gauge transformations (Einstein called them \qq{$\la$-transformations} \cite{ein1955-2}) remains:
\disn{s20}{
{\Gamma'}^\alpha_{\mu\nu} = {\Gamma}^\alpha_{\mu\nu} + (\dd_\mu \la) \delta^\alpha_\nu.
\nom}
The expression \eqref{s18} is strictly invariant with respect to \eqref{s20} (boundary terms do not appear) only in the additional assumption $\la \str{x^0\to\pm\infty} 0$ analogous to \eqref{s19} which again does not affect the invariance of the field equations.
One can notice that simultaneous coordinate change and transformation \eqref{s20}  does not lead to the appearing of boundary terms if the condition
\disn{s21}{
\ln\det\left| \frac{\partial x'}{\partial x} \right|+4\la\str{x^0\to\pm\infty} 0.
\nom}
is satisfied. The Ricci tensor turns out to be invariant with respect to \eqref{s20} (but not to \eqref{gauge}), see \eqref{ss1}.

The theory corresponding to the choice of $S_1+S_2+S_3$ as a final form of action turns out to be self-inconsistent. Indeed, since the increment \eqref{s18} which was added to the original action \eqref{1} partially broke the gauge invariance with respect to \eqref{s20}, three of four degrees of freedom of $A_\m$ (see \eqref{s12}), which initially were gauge ones, become physical ones. However, they did not become dynamical ones since $S_1$ and $S_2$ do not depend on them and $S_3$ depends only linearly. Varying with respect to them leads to the equations
\begin{align}\label{s22}
	\dot{x}_j^\be(\ta) = 0,
\end{align}
which are self-inconsistent because correct parametrization of the worldline means that this quantity cannot vanish. One can solve this problem by choosing the more complicated expression than EH term \eqref{1} for the action of gravity, so abovementioned degrees of freedom may become dynamical ones. In choosing such an expression one should keep in mind the following fact.

First of all, we note that in the framework of GR, when the only independent variable is metric, whereas connection is uniquely expressed through it as Christoffel symbols \eqref{s5}, the EH action is the only generally covariant action which is not lead to the higher derivatives in the field equations. However, for the theory considered here, in which the connection is independent, it is possible to construct some other invariants which possess such a property. To construct them one can use the curvature tensor \eqref{s1} as well as the torsion tensor \eqref{s3}.
In contrast with the case of Riemannian geometry corresponding to GR, when only one nontrivial contraction of curvature tensor exists (the one that leads to the Ricci tensor \eqref{s2.1}), in the case of independent connection there is one more nontrivial contraction:
\disn{s23}{
R^\m{}_{\m\al\be}=\dd_\al\Gamma^\m_{\be\m}-\dd_\be\Gamma^\m_{\al\m}.
\nom}
This antisymmetric tensor was introduced early in the Eddington theory (see below in the section \ref{istor}). According to Cartan's terminology it is called a \textit{segmentary curvature} or a \textit{homothety curvature}, see \cite{vizgin}. Let us choose a new term in addition to EH one \eqref{1} as an action of gravity in the form of quadratic expression with respect to contraction \eqref{s23}:
\disn{s24}{
S_4 = -\theta \int\! d^4 x\, \sqrt{-g}\,g^{\al\ga}g^{\be\de}
 R^{\mu}{}_{\mu\alpha\beta} R^{\nu}{}_{\nu\ga\de},
\nom}
where $\te$ is an arbitrary constant.

The resulting form of the action for the considered theory is the following:
\disn{s25}{
S = (S_1+S_4)+S_2+S_3=
-\int\! d^4 x \,\sqrt{-g}\, \left(\frac{1}{2\ka}g^{\nu\be}R^\m{}_{\n\mu\be}+
\theta g^{\al\ga}g^{\be\de}
R^{\mu}{}_{\mu\alpha\beta} R^{\nu}{}_{\nu\ga\de}\right)-\ns-
\sum_j\int\! d\ta \ls m_j \sqrt{\dot{x}^{\mu}_j(\ta)\dot{x}^\nu_j(\ta) g_{\mu\nu}(x_j(\ta))} + \frac{q_j}{4\om} \dot{x}^{\be}_j(\ta)\Gamma^\m_{\be\m}(x_j(\ta))\rs.
\nom}
It is invariant with respect to diffeomorphisms as well as to gauge transformations \eqref{s20} restricted by the condition \eqref{s21}. To obtain the full set of field equations one need to vary this action with respect to all independent variables describing gravity ($g_{\m\n}(x)$ and $\Gamma^\al_{\m\n}(x)$) as well as the particles ($x^\m_j(\ta)$).

\section{The field equations}\label{urdv}
Firstly let us find the variation of the action \eqref{s25} with respect to the connection $\Gamma^\al_{\m\n}(x)$. Such a variation of the $S_1$ was already calculated in \eqref{s8}, whereas the term $S_2$ does not give a contribution. For $S_4$, assuming \eqref{s23}, we have
\disn{s26}{
\de S_4=4\te\int\! d^4 x \,\sqrt{-g}\,\ls \bar D_\al R^\n{}_\n{}^{\al\be}\rs \de\Gamma^\m_{\be\m},
\nom}
where $\bar D_\al$ is a Riemannian covariant derivative with the connection \eqref{s5}.
For the remaining term $S_3$ we obtain
\disn{s27}{
\de S_3=-\sum_j \frac{q_j}{4\om}\int\! d\ta\, \dot{x}^{\mu}_j\,\de\Gamma^\m_{\be\m}(x_j)=
-\frac{1}{4\om}\int\! d^4 x \,\sqrt{-g}\,j^\be \de\Gamma^\m_{\be\m},
\nom}
where we have introduced the quantity
\disn{s28}{
j^\m=\sum_j q_j \int\! d\ta\,\dot x^\m_j \de(x-x_j)\frac{1}{\sqrt{-g}}
\nom}
which is no other than four-current of relativistic particles if one interprets $q_j$ as their electric charges.

As a result we obtain a field equation which replaces \eqref{s9}:
\disn{s29}{
\frac{1}{2\ka}\ls C^{\m\n\be}+C^{\n\be\m}-C_\ga{}^{\ga\m} g^{\n\be}-C_\al{}^{\n\al}g^{\be\m}\rs+
\ls 4\te\bar D_\al R^\ga{}_\ga{}^{\al\be}-\frac{1}{4\om}j^\be\rs g^{\m\n}=0.
\nom}
Multiplying it on $g_{\m\n}$, one easily notices that it leads to the satisfaction of \eqref{s9} (and therefore \eqref{s12}) and
\disn{s30}{
4\te\bar D_\al R^\ga{}_\ga{}^{\al\be}-\frac{1}{4\om}j^\be=0.
\nom}
Substituting \eqref{s12} in \eqref{s23} and making use of \eqref{s31}, it is easy to notice that
\disn{s31.1}{
R^\ga{}_{\ga\al\be}=4\om\ls \dd_\al A_\be-\dd_\be A_\al \rs.
\nom}
As a result one can rewrite \eqref{s30} in the following form:
\disn{s32}{
\bar D_\al F^{\al\be}=4\pi j^\be,
\nom}
where
\disn{s33}{
F_{\al\be}=\dd_\al A_\be-\dd_\be A_\al,
\nom}
and the arbitrary constant $\om$ which was introduced above \eqref{s10} had now been fixed:
\disn{s33.1}{
\om=\frac{1}{16\sqrt{\pi\te}}.
\nom}
So the varying of action \eqref{s25} with respect to the connection leads to the expression for the connection \eqref{s12} as well as the equation \eqref{s32} which reproduces Maxwell equation in curved spacetime if one identifies the part $A_\m$ of connection DoFs with the EM potential. Note that under gauge transformations \eqref{s20}, which do not change the full action \eqref{s25}, the quantity  $A_\m$ transforms as follows (according to \eqref{s12}):
\disn{s33.2}{
A'_\m=A_\m+\frac{1}{\om}\dd_\m\la,
\nom}
as EM potential should transform.

Now let us vary the action \eqref{s25} with respect to the particles
coordinates $x^\m_j(\ta)$, so contributions will be given only by $S_2$ and $S_3$ contributions.
They have the well-known form of the action of relativistic particles and action of its interaction with the EM potential of the form $\Gamma^\m_{\be\m}/(4\om)$ (if $q_j$ are charges).
So one can immediately write the known equations of motion appearing after the variation with respect to $x^\m_j(\ta)$:
\disn{s34}{
m_j u^\n_j \bar D_\n u_{j\al}=q_j F_{\al\n} u^\n_j,
\nom}
where we noticed that making use of \eqref{s12}, \eqref{s31} and \eqref{s33} leads to the equality
\disn{s35}{
\dd_\al \ls\frac{1}{4\om}\Gamma^\m_{\n\m}\rs-\dd_\n\ls\frac{1}{4\om}\Gamma^\m_{\al\m}\rs=\dd_\al A_\n-\dd_\n A_\al=F_{\al\n}.
\nom}
In the equation \eqref{s34} we have introduced a normalized 4-velocity vector of the relativistic particle
\disn{s36}{
u^\al_j=\frac{\dot x^\al_j}{\sqrt{\dot{x}^{\mu}_j\dot{x}^\nu_j g_{\mu\nu}(x_j)}}.
\nom}
The resulting expression \eqref{s34} obviously reproduces the equations of motion of relativistic particle in the gravitational field with metric $g_{\m\n}$ and EM field defined by the certain part of the connection treated as EM potential $A_\m$.

Let us finally vary the action \eqref{s25} with respect to metric $g_{\m\n}$. Contributions will be given by $S_1, S_2$ and $S_4$. The contribution of $S_1$ was calculated in
the section~\ref{urav} (see the left-hand side of \eqref{s16}). Merging it with contribution of $S_4$
(which can be easily calculated), we obtain
\disn{s37}{
\de (S_1+S_4)=\int\! d^4 x \,\sqrt{-g}\,\ls
\frac{1}{2\ka}\bar G^{\al\be}+2\te R^{\mu}{}_\mu{}^{\alpha\ga} R^{\nu}{}_\nu{}^\be{}_\ga-
\frac{\te}{2}g^{\al\be}R^{\mu}{}_\mu{}^{\ga\de} R^{\nu}{}_{\nu\ga\de}
\rs\de g_{\al\be}.
\nom}
The contribution of $S_2$
\disn{s38}{
\de S_2=-\sum_j \frac{m_j}{2} \int\! d\ta\, \sqrt{\dot{x}^{\mu}_j\dot{x}^\nu_j g_{\mu\nu}(x_j)}u^\al u^\be \de g_{\al\be}(x_j)=
-\frac{1}{2}\int\! d^4 x \,\sqrt{-g}\, T_\text{p}^{\al\be}\de g_{\al\be}
\nom}
is determined by the energy-momentum tensor of the relativistic particles
\disn{s39}{
T_\text{p}^{\al\be}=\sum_j m_j \int\! d\ta\, \sqrt{\dot{x}^{\mu}_j\dot{x}^\nu_j g_{\mu\nu}(x_j)}\de(x-x_j)\frac{1}{\sqrt{-g}}u^\al u^\be.
\nom}
As a result we obtain the field equations in the following form:
\disn{s40}{
\bar G^{\al\be}=\ka\ls T_\text{p}^{\al\be}-4\te\ls
R^{\mu}{}_\mu{}^{\alpha\ga} R^{\nu}{}_\nu{}^\be{}_\ga-\frac{1}{4}g^{\al\be}R^{\mu}{}_\mu{}^{\ga\de} R^{\nu}{}_{\nu\ga\de}\rs\rs.
\nom}
Using the corollary \eqref{s31.1} of the other field equations, the denotation \eqref{s33} and choosing of constant $\om$
\eqref{s33.1}, one easily notices that this equation is the Einstein one:
\disn{s41}{
\bar G^{\al\be}=\ka\ls T_\text{p}^{\al\be}+T_\text{EM}^{\al\be}\rs,
\nom}
where
\disn{s42}{
T_\text{EM}^{\al\be}=-\frac{1}{4\pi}\ls
F^{\alpha\ga} F^\be{}_\ga-\frac{1}{4}g^{\al\be}F^{\ga\de} F_{\ga\de}\rs
\nom}
reproduces the expression for the energy-momentum tensor of EM field with potential $A_\m$.

Therefore the full set of field equations corresponding to the action \eqref{s25} turns out to be the set of the equations \eqref{s32},\eqref{s34}, and \eqref{s41} together with the expression for the connection \eqref{s12}. These equations reproduce the Einstein-Maxwell equations if one identifies $A_\m$ in \eqref{s12} with the EM potential.

Such a natural unification of gravitation and electromagnetism appears if we restrict ourselves to the consideration of the matter in the form of classical relativistic point particles.
It is quite simple to generalize this consideration on the case of continuous media, whereas the generalization on the important case of matter fields (i.~e. spinor or even scalar electrodynamics) is a very non-trivial problem.
We will discuss it in the concluding section~\ref{zakl}. Now let us shortly review the ideas of classical UFTs related to the one that we described above.

\section{Historical notes}\label{istor}
The present section is devoted to the short discussion of the known UFTs in comparison with the approach described above. We do not consider extra dimensions because such theories are too much different from the above one. The main characteristics of the above theory which can be different in other theories are:

\begin{itemize}
\item the symmetricity of metric,
\item the non-symmetricity of connection (the presence of torsion),
\item the initial independence of metric and connection.
\end{itemize}
The literature concerning unified field theories is quite wide. We do not pretend to give a full comprehensive review here. Such reviews can be found, e.~g. in the monographs \cite{vizgin}, \cite{tonnelat} as well as in the reviews \cite{goenner1,goenner2}.

The most famous of the early attempts of the unification of gravity with electromagnetism is assumed to be Weyl theory which was firstly proposed in 1918 in \cite{weyl_1918} (see also \cite{weyl_raum}). The main idea of the Weyl approach is an introduction of an additional gauge invariance to the theory, namely the scaling invariance of metric (which is symmetric)
\begin{align}\label{is1}
{g'}_{\mu\nu}(x) = e^{2\lambda(x)} g_{\mu\nu}(x).
\end{align}
These transformations are often called conformal, which is not quite correct, since they (in contrast with conformal ones) are not a part of diffeomorphisms group and were postulated by Weyl besides of it. It is more convenient to call them \textit{Weyl transformations}.
The adjusting of the parallel transport rule with this new symmetry leads to the appearance of additional degrees of freedom $A_\m$ in the connection, which in Weyl theory takes the form
\disn{is2}{
\Gamma^\al_{\m\n}=\bar \Gamma^\al_{\m\n}+\frac{1}{2}\ls
A_\m\de^\al_{\n}+A_\n\de^\al_{\m}-A_\be g^{\be\al}g_{\m\n}\rs,
\nom}
(cf. \eqref{s12}), where $\bar \Gamma^\al_{\m\n}$ is, as above, the Riemannian connection \eqref{s5}. Note that in Weyl theory as well as in GR the connection is symmetric, in contrast with the above theory.
Under the gauge transformation \eqref{is1} additional DoFs are transformed as follows:
\disn{is3}{
A'_\be=A_\be-\dd_\be\ln\la,
\nom}
which allowed Weyl to interpret this quantity as an EM potential. The curvature tensor in Weyl theory, as well as in the above approach, splits into two terms:
\disn{is4}{
R^\m{}_{\n\al\be}=\bar R^\m{}_{\n\al\be}+\frac{1}{2}\de^\m_\n\ls\partial_\al A_\be - \partial_\be A_\al\rs,
\nom}
cf. \eqref{s14.1}.

However, the gauge invariance with respect to \eqref{is1} postulated by Weyl entailed two undesirable consequences. Firstly, the square of interval which was observable in GR, is no longer observable: \qq{only the ratios of the metric components have direct physical meaning}, and not these quantities themselves \cite{vizgin}. As a result, some additional \textit{ad hoc} methods of length measurements are required, see \cite{vizgin}. Secondly, the condition of Weyl-invariance of the action of the theory does not allow to use EH action \eqref{1} in it since EH action is not invariant with respect to \eqref{is1}. Instead, one has to use squared curvature tensor which leads to the poor agreement with the observables.

Another attempt to unify gravity with electromagnetism was made by Eddington \cite{eddington} in 1921. If we consider the Weyl approach as less \qq{strict} than GR one, since Weyl did not require the length conservation under a parallel transport while keeping the conservation of angles, than the Eddington approach is much less strict: he did not require the angles conservation as well, assuming that the connection is not related to metric at all. In this sense his approach is close to above one, but there is one essential difference: Eddington assumed the connection to be symmetric (as in the Weyl theory and GR), so in this sense his approach coincides with the Palatini one. In Eddington theory antisymmetric part of the Ricci tensor (cf. \eqref{s15}) and segmentary curvature \eqref{s23} do appear; due to the symmetricity of the connection they are related by the condition
\disn{is5}{
R_{\al\be}-R_{\be\al}=R^\m{}_{\m\al\be}=\dd_\al\Gamma^\m_{\be\m}-\dd_\be\Gamma^\m_{\al\m}.
\nom}
Note that from \eqref{s15} and \eqref{s31.1} one can see that in the above approach a similar but not quite equivalent condition is satisfied: antisymmetric part of the Ricci tensor is doubled.

The trace of the connection in the Eddington theory is identified with the EM potential (as in the section \ref{urdv}, see below \eqref{s33.1}), so the antisymmetric part of the Ricci tensor turns out to be the EM field tensor. The symmetric part of Ricci tensor, according to Eddington, is related to the metric by the condition (which is postulated without any variational principle)
\disn{is6}{
R_{\al\be}+R_{\be\al}=\La g_{\al\be},
\nom}
where $\La$ is a certain universal constant. The condition \eqref{is6} is called by Eddington \qq{natural world gauge}. The first and foremost significant drawback of the Eddington theory is the fact that field equations (including Einstein ones) are not derived from a variational principle, but instead are postulated using \qq{the \q{principle of identification} of the elements of the geometrical structure  developed by him with fundamental physical quantities}, see~\cite{vizgin}.

It seems that the non-symmetric connection was firstly introduced by Schouten and Friedman in 1923-24 \cite{schouten,schouten_fr}. They suggest to consider \textit{half-symmetric} connection, which antisymmetric part has the form
\disn{is7}{
\Gamma^\al_{\m\n}-\Gamma^\al_{\n\m}=A_\m\de^\al_{\n}-A_\n\de^\al_{\m},
\nom}
whereas the Lagrangian of the theory is chosen to be the square root of Ricci tensor determinant (the simplest scalar density which can be constructed only from connection).
A similar idea was used much later in 1982 in \cite{ferraris1982} when the authors wrote the action of unified theory of gravitation and electromagnetism. However, they used only the symmetric part of the Ricci tensor and the whole action was far more complicated. Note that in this paper the role of the EM potential was played by the same components of the connection that in the above theory: the field equations entailed the satisfaction of the relation \eqref{s12} for the connection.

Needless to say, the man who most wanted to find the unified field theory was Einstein. One of the directions of his quest was quite close to the above approach. The difference between them is the Einstein's assumption of non-symmetricity of
metric\footnote{According to Goenner \cite{goenner1}, this idea was suggested to him by R.~F$\ddot{\text{o}}$rster, who thus repelled Einstein from the one of the most natural formulations of the unified theory.}.
Einstein had been working on such a generalization of GR since 1917 (see \cite{goenner1}); he wrote the first paper \cite{ein_conn} on this topic in 1925, whereas the last one \cite{ein1955-2} was written by him in 1955 shortly before his death and turned out to be his last paper ever.

In the approach developed by Einstein the action was chosen to be EH one \eqref{1} in which the metric and the connection are independent. Initially they are completely arbitrary, so the Einstein approach differs from the one that was described in section \ref{urav} only by the non-symmetricity of metric. As it was mentioned in the Introduction, Einstein in fact proposed the method which (for symmetric metric and connection) later became known as Palatini formalism.

It was shown in \cite{ein_conn} that among the appearing field equations a part of vacuum (i.~e. without current) Maxwell equations is present if one identifies the antisymmetric part of the metric with the EM field
tensor\footnote{Note that in the papers \cite{ein1955-1,ein1955-2} such an identification was no longer discussed, while suggested modification of GR is treated as \qq{an attempt to construct a theory of the total field by generalizing the equations of the purely gravitational field} rather than unification of gravity and electromagnetism.}.
The remaining equations are initially analyzed in the absence of EM field, and the manual elimination of torsion is performed to reproduce GR. For the general case Einstein performed an approximate analysis of the appearing equations, but he failed to construct a satisfactory UFT. The attempts of developing of theories with non-symmetric metrics had been made lately, e.~g. in \cite{Borchsenius1976} where it was noticed that in such a framework it is possible to identify the trace of the torsion with the EM potential (as in the section \ref{urav}, see~\eqref{s14}), and the action contained the interaction of this quantity with the current density.

Although Einstein paid a special attention to gradient transformations \eqref{s20} which he called \qq{$\la$-transformations} \cite{ein1955-2}, he did not connect them to gradient transformations of the EM potential (this occurred quite naturally in the above approach, see~\eqref{s33.2}). In the Einstein theory the action turns out to be invariant with respect to \eqref{s20} due to invariance of curvature tensor, but not with respect to more general transformations \eqref{gauge} since the metric is not symmetric and, according to \eqref{ss1}, its contraction with Ricci tensor remains dependent on the transformation parameter $f_\m$. If the metric is symmetric, then EH action is invariant with respect to \eqref{gauge} as well, which is discussed in the section \ref{urav}. The restriction of symmetry to \eqref{gauge} in the above approach happens only after the addition of the term \eqref{s18} in the action, which couples the connection to the matter.

Einstein was close to above approach once again: in the several papers, first of which was the paper \cite{ein_1923} written in 1923. In these papers he considered a certain implicit action constructed from symmetric and antisymmetric part of the Ricci tensor. It corresponds to the action of gravity $S_1+S_4$ that was used in the section \ref{mater}, in which the part $S_1$ is constructed from symmetric part of the Ricci tensor (in these papers Einstein assumed that both the metric and the connection is symmetric), while the part $S_4$ can be treated as constructed from antisymmetric one due to its coincidence \eqref{is5} with the segmentary curvature in the theory with symmetric connection. However, the condition of connection symmetricity made this UFT (which was eventually abandoned by Einstein) drastically different from the above one.

The approach proposed by Einstein in \cite{ein_1923} was generalized on the case of non-symmetric connection by young French mathematician Henri Eyraud in 1925\cite{eyraud}. Following Einstein, he assumed that the action somehow depends on symmetric and antisymmetric part of Ricci tensor, but in the presence of torsion the condition \eqref{is5} is no longer holding, so he failed to obtain the explicit form of the action ($S_1+S_4$) as well as Einstein himself. However, Eyraud obtained an expression for the non-symmetric connection in the form of  \eqref{s12}, while a part of its degrees of freedom $A_\m$, which (considering \eqref{s31}) is the same that the trace of the connection $\Gamma^\al_{\m\al}$ up to the gradient transformation and a constant multiplier, was interpreted as the EM potential. The same expressions and interpretations were used in the paper by Straneo in 1931, although the expression \eqref{s12} was postulatively introduced there rather than derived from any variational principle.

Proceeding from the analyzed publications, we conclude that, strangely enough, the approach described in the section \ref{urav}-\ref{urdv} had not been discovered in the first half of XX century, at the time of the most intense search for UFT.

First paper in which the variant of such an approach was formulated is the 1978 paper \cite{Krechet}. However, it was published only in Russian in an obscure journal and thus has not been cited in the similar works. In \cite{Krechet} the authors start from EH action with symmetric metric and independent non-symmetric connection. Then they discover that the trace of the torsion (in fact it is the same that the part of the DoFs of connection corresponding to $A_\m$ in \eqref{s12}, see~\eqref{s14}) remains arbitrary \textit{on-shell}. Then they propose to supplement the action by the squared segmentary curvature $S_4$ \eqref{s24} and discover the coincidence of resulting expressions with the Einstein-Maxwell ones if the trace of the \textit{torsion} is identified with EM potential. In particular, they discover the possibility of interpretation of \qq{$\la$-transformations} (which are the symmetry transformations of the resulting theory) as gauge transformations of EM potential. Then the authors obtain the equations of motion of test particles \eqref{s34}, although they derive it from the \textit{ad hoc} condition (\qq{physical trajectories in the space-time} must have a least length among those along which the length of vector by parallel transport is conserved) rather than from the action, as it was done in section~\ref{urdv}. The particles is considered precisely as test ones, so backreaction (the appearing of the corresponding contributions in equations of motion \eqref{s32} and \eqref{s41}) is not discussed.

Almost simultaneously with the paper \cite{Krechet} (also in 1978) the Palatini formulation with non-symmetric connection was studied in the paper \cite{hehl}. However, the authors of \cite{hehl} did not make an attempt to use obtained result in the construction of UFT. Instead of it they studied the relation of the appearing theory (squared segmentary curvature was not added to the action) to the usual description of gravity in the framework of Riemannian geometry. This studies have been continued in the recent papers \cite{dadhich2012,bernal2017} which indicates a continuing interest to the idea of using non-symmetric connection. The possibility of addition of squared segmentary curvature to obtain a kind of Einstein-Maxwell theory that proposed in \cite{Krechet} was rediscovered independently in the paper \cite{tucker1995}. In the framework of the resulting theory the authors also considered a spherically symmetric solution corresponding to Reissner-Nordstr$\ddot{\text{o}}$m one.

This concludes our historical survey of the papers devoted to UFT which are related to the approach described in the sections \ref{urav}-\ref{urdv}.

\section{Conclusion}\label{zakl}
In the present paper we described a possible way of unification of gravity and electromagnetism, which had been somehow missed at the time of the most intensive search for such an unification in the first half of XX century. The idea of such way was proposed only in 1978 in the paper \cite{Krechet}.

It must be noted that in this framework the matter interacting with EM field can be defined only in \textit{purely classical sense}: as a set of point particles. The generalization on the case of ideal fluid (i.~e. continuous medium consisting on classical particles) is quite simple. For the description of such a matter one can choose a current $J^\m(x)$ as an independent variable, which is a vector density and satisfies the continuity equation $\dd_\m J^\m=0$. Various forms of the action of the ideal fluid based on this choice are examined in \cite{statja48}. The part of the action $S_3$ describing the interaction of the ideal fluid with the connection  $\Gamma^\m_{\be\al}$ can be written by analogy with \eqref{s18} in the following form:
\disn{zz1}{
S_3=-\frac{q}{4\om}\int\! d^4 x\,J^\be\Gamma^\m_{\be\m},
\nom}
while the current $j^\m$ appearing in the equations of motion \eqref{s32} takes the form
\disn{zz2}{
j^\m=\frac{q}{\sqrt{-g}}J^\m
\nom}
instead of  \eqref{s28}. In such a simple case all the matter has fixed (equal to $q$) charge-to-mass ratio, although one could introduce multiple types of matter with different values of such a ratio.

However, from the modern point of view the purely classical description of matter is obviously unsatisfactory as in the quantum theory the EM field interacts with complex (usually spinorial) matter fields rather than particles or continuous media. The EM potential ${\cal A}_\m$ turns out to be Abelian gauge field corresponding to local group of inner symmetry $U(1)$, and takes part in the definition of covariant derivative corresponding to that symmetry:
\disn{z1}{
\na_\m\psi=(\dd_\m+i {\cal A}_\m)\psi,
\nom}
i.~e. it multiplies on the imaginary unit. The lack of imaginary unit at the $A_\m$ in the expression for the connection \eqref{s12}
(assuming that the connection is real) means that $A_\m$ should be treated  as a gauge field corresponding not to local phase transformations $U(1)$
\disn{z2}{
\psi'(x)=e^{-i\la(x)}\psi(x),
\nom}
but rather to local \emph{scale} transformations of tensorial matter
fields\footnote{
Note that it is not the presence of complexity itself in \eqref{z1} (which can be avoided by choosing $SO(2)$ as a real form of $U(1)$) is essential, but rather the fact that the quantity which \qq{elongates} the usual derivative is not reduced to simple scaling of matter fields as it occurs for scale transformations \eqref{z3}.}
\disn{z3}{
\psi'^\m(x)=e^{-\la(x)}\psi^\m(x),
\nom}
which are analogous to the scale transformations of the Weyl theory mentioned in the section ~\ref{istor} (see \eqref{is1}). A \qq{charge} of matter field in this case equals to difference between the number of its upper and lower indices (and is thus quantized \textit{on the classical level}).

Therefore the interpretation of $A_\m$ which presents in the real-valued expression \eqref{s12} as an EM potential in the field theory is possible if one somehow provides the appearance of imaginary unit at it. The efforts of researchers working on such an approach were mainly concentrated on the complexification of the connection to provide its interaction to the matter fields (see \cite{horie9409018th}, \cite{horie2} and the references therein). The consideration was usually performed in terms of frame bundles. It must be stressed that if the matter does not described in terms of fields, there is nowhere from which the global and certainly local $U(1)$ symmetry can arise since in the Lagrangian of classical electrodynamics with matter in the form of point particles there is no variables that can be affected by such transformations.

The question whether it is possible to obtain a satisfactory UFT in the framework of the above approach in the field-theoretic description of matter remains open. Nevertheless, the classical electrodynamics of charged particles appears in such a framework exceptionally simple, even simpler than in the most famous UFT, namely the Kaluza-Klein one \cite{wesson_kaluza}. In the Kaluza-Klein theory one needs to make a physical proposition about the existence of fifth dimension to obtain \qq{electromagnetic} degrees of freedom, whereas the above approach naturally appears when one \emph{eliminates} the postulate of the absence of torsion in the GR. This fact allows to put this approach on a par with the most famous classical UFTs and, in our opinion, assigns the sufficient historical and methodical value to it.

{\bf Acknowledgements.}
The work was supported by SPbSU grant N~11.38.223.2015.
One of the authors (A.~S.) thanks the Library of Russian Academy of Sciences for the assistance in working with sources.

\end{document}